\documentclass[%
 aip,
 amsmath,amssymb,
 reprint,%
]{revtex4-1}

\usepackage{graphicx}% Include figure files
\usepackage{dcolumn}% Align table columns on decimal point
\usepackage{bm}% bold math
\usepackage[utf8]{inputenc}
\usepackage[T1]{fontenc}
\usepackage{mathptmx}
\usepackage{etoolbox}

\makeatletter
\def\@email#1#2{%
 \endgroup
 \patchcmd{\titleblock@produce}
  {\frontmatter@RRAPformat}
  {\frontmatter@RRAPformat{\produce@RRAP{*#1\href{mailto:#2}{#2}}}\frontmatter@RRAPformat}
  {}{}
}%
\makeatother
\begin{document}

\preprint{AIP/123-QED}

\title{Freezing of the tetrahedral amorphous network in supercooled water triggers crystallization towards LDA ice}
% Force line breaks with \\
\author{Ashutosh Srivastava}
\author{Pankaj A. Apte}%
 \email{papte@iitk.ac.in}
\affiliation{ 
Department of Chemical Engineering, Indian Institute of Technology Kanpur, Kanpur, UP 208016, India%\\This line break forced with \textbackslash\textbackslash
}%

\date{\today}% It is always \today, today,

\begin{abstract}
In this work, we provide mechanistic insight into the initial stages of formation of ice across the limit of stability of supercooled water.   Such an analysis is particularly important since crystal nucleation is not a relevant mechanism under these conditions.   Using molecular dynamics simulation with the TIP4P/2005 potential, water is cooled at a constant pressure with  cooling  rates of 5 to 10 K per nanosecond.    As the liquid is cooled across the temperature of maximum density ($\mbox{T}_0 = 277$ K), we find that there is a continuous increase in the tetrahedrality of the system.   As the cooling continues across the limit of stability of water ($\mbox{T}_s \approx 235$ K), large scale thermal fluctuations dissipate while the thermal equilibration is achieved through small scale fluctuations.  This phenomenon, known as the {\it dynamical crossover} [Goutam et. al. in J. Stat. Phys., 168: 1302--1318 (2017)],  ends the existence of the liquid state.   Subsequently, we find that the tetrahedral network drives the  decrease of energy and density.   This process terminates when the network undergoes `freezing' (i.e., the bonds of the network acquire sufficient rigidity), due to which the network evolution, as a whole, stops.  This triggers a qualitative change in the relaxation mechanism: subsequent relaxation occurs through crystallization, i.e., an increase in the structural order.   In particular, we find that the cubic and hexagonal crystalline motifs, which possess medium range order, increase rapidly across the freezing point. In the resulting LDA ice states, cubic ice is found to have a significant contribution in the overall extent of crystallization, which is consistent with the experimental findings.  Overall, our work provides the specific mechanism by which crystallization (leading to LDA ice) is initiated  across the limit of stability of supercooled water. 
\end{abstract}

\maketitle

%%%%%%%%%%%%%%%%%%%%%%%%%%%%%%%%%%%%%%%%%%%%%%%%%%%%%%%%%%%%%%%%%%%%%%%%%%%%%%%%%%%%%%%%%%%%%%%%%%%%%%%%%%
\section{\label{sec:level1} Introduction}

Understanding the mechanism of initial stages of ice formation in deeply supercooled water is of potential importance to processes that occur naturally in extremely cold climates as well as in technological applications such as cryopreservation of tissues.   Across the limit of stability, the supercooled liquid water transitions into low density amorphous (LDA) ice.  Experimental studies have revealed that LDA ice possesses tetrahedral network with structural features at the intermediate or medium range order (MRO) level~\cite{BENMORE05,ELLIOTT95,MAYER87,MURRAY06,MALKIN12,MALKIN15} with a significant percentage of cubic ice.~\cite{AMAYA17}   The experimentally deduced structure of LDA ice~\cite{MALKIN12,MALKIN15} suggests that the development of MRO order occurs due to aggregation of polyhedra or vitrites corresponding to both cubic and hexagonal ices, as originally predicted by Stillinger.~\cite{STILLINGER80}    Due to fast crystallization in the deeply supercooled state, these structural changes cannot be studied experimentally and therefore, simulation methods have been used extensively to analyze such processes.       
Using simulations with TIP4P model of water, Matsumoto et. al.~\cite{mMATSUMOTO02} investigated the freezing of water in a supercooled state at 230 K and found that the fast freezing of water is accompanied by formation of hydrogen bonded (tetrahedral) network consisting of six-membered rings  that spans the entire network.  This shows the build-up of MRO structures involves six membered rings (that constitute hexagonal and cubic ice motifs).  In another simulation study using TIP4P model, Malkin et. al.~\cite{MALKIN12} found that during crystallization at 220 K (which is below the melting temperature of 232 K for TIP4P) stacking disorder emerges with about 50 \% content each of cubic and hexagonal ices.   In investigation using monatomic (mW) water model,~\cite{MOLINERO09} Molinero and co-workers concluded that a rapid increase in the 4-coordinated molecules with supercooling of water leads to a sharp yet continuous relaxation to LDA  ice at the limit of stability ($\mbox{T}_s \approx 205$ K in case of mW water).~\cite{MOORE11-1, HUJO11, MOORE10}   The degree of crystallization in LDA ice at 150 K, was found to be around 5 \% with a significant percentage of interfacial water.~\cite{MOORE11-2}   Thus different simulation methods have predicted the development of MRO structures during relaxation of supercooled water which leads to LDA ice.  However, at or close to the limit of stability,  nucleation is not a relevant mechanism.  Thus a key issue,  which has not been addressed, is the mechanism by which crystallization is initiated in the deeply supercooled water. Another aspect of the transition which also needs clarity is the role of the hydrogen bonded  tetrahedral (amorphous) network.  Due to inherent cooperativity of the hydrogen bonds,~\cite{STILLINGER80} tetrahedral network is expected to contribute significantly to the overall changes of energy and density.    In an investigation using TIP4P/2005 water model, Matsumoto~\cite{mMATSUMOTO09} found that the volume expansion of deeply supercooled water is directly related to increase in the extent of the tetrahedral network.   Since formation of LDA ice from supercooled water also involves volume expansion, the changes associated with the tetrahedral network (apart from the structural changes) may also be important in the transition.   

In this work, we study the transition of  supercooled water to LDA ice  using simulation with TIP4P/2005 potential model with a focus on the two aspects (discussed above): (i) the detailed mechanism through which structural changes (leading to LDA ice) are initiated and (ii) the role of tetrahedral network in the transition.   
An important issue is the presence of thermal fluctuations in the supercooled liquid state which tend to disrupt any structural motifs that are formed.  Recently, using simulations with monatomic (mW) potential of water, it was shown that the transition towards LDA ice is initiated with a {\it dynamical crossover} in which large scale thermal fluctuations dissipate, ending the existence of the liquid state.~\cite{GAUTAM17} This indicates that the dynamical crossover preludes the development of structure leading to LDA ice.  In this work, we show that the same mechanism applies to water simulated with TIP4P/2005 potential.    Our results also indicate that there exists a `transition region' in which tetrahedral network drives the relaxation (following the dynamical crossover). The transition region ends  with the `freezing' of the network implying that the bonds in the network develops sufficient rigidity.  The freezing of the network triggers crystallization (structural ordering) towards LDA ice.     Below, we describe our simulation methodology followed by the detailed results and analysis of the changes during crystallization of supercooled water.

\section{Methodology}

Molecular dynamics simulations were performed in the isothermal-isobaric NPT ensemble with the LAMMPS package~\cite{PLIMPTON95} using Nos\'{e}--Hoover  thermostat and barostat. The
thermostat damping constant was 200 fs and the barostat damping constant was 1000 fs. The equations of motion were integrated using the velocity-Verlet scheme with a timestep
of 1 fs.  A system size of 27000 molecules was used with periodic boundary conditions on all 3 dimensions to maintain bulk behavior.  A constant pressure of 1 atm was maintained while the temperature was decreased linearly at rates of 5 and 10 K/ns. Long-range electrostatic interactions were handled using PPPM for with a target accuracy of $10^{-4}$. Tail corrections were enabled for the Lennard-Jones 
term to compensate for truncation effects. In
addition, the SHAKE constraint algorithm (with a tolerance of $10^{-4}$) was applied to all water molecules to maintain rigid molecular geometry during the dynamics. 

The hydrogen bonds were identified with a geometric criteria:\cite{LUZAR93,LUZAR96} two water molecules are considered to be bonded if the oxygen-oxygen distance is less than 3.5 A$^0$ and the angle between the oxygen--oxygen axis and one of the O-H bonds less than 30$^0$.
Using these criteria, the largest connected network of 4-coordinated water molecules (henceforth referred to as the tetrahedral network) was identified and
the fraction of molecules in the network (denoted as f4c) was computed.   To measure the deviation from the ideal tetrahedral angle, we define the average angular deviation (AAD) as follows:

\begin{equation}
\mbox{AAD} = \frac{1}{\mbox N}{\sum_{i=1}^{\mbox N}}{\mbox{AD}_i}
\label{eq:aad}
\end{equation}

where, AD$_i$ is the angular deviation of the i$^{\mbox{\small th}}$ molecule which is defined by the
following equation : 

\begin{equation}
\mbox{AD}_i = {\sum_{j=1}^{\mbox{\small nn}_i-1}}{\sum_{k=j+1}^{\mbox{\small nn}_i}}{\left(\mbox{cos}~{\theta}_{jik} + \frac{1}{3}\right)^2},
\label{eq:adi}
\end{equation}
where $\mbox{\small nn}_i$ is the number of nearest neighbors of i$^{\mbox{\small th}}$ molecule and $\theta_{jik}$ is the angle subtended by the vectors $\vec{r}_{ij}$ and $\vec{r}_{ik}$  at the i$^{\mbox{\small th}}$ molecule.  When computing the AAD of the entire system, we consider N in Eq.~(\ref{eq:aad}) as the number of molecules in the entire system except the molecules with less than 2 nearest neighbors since angular deviation is not relevant for such molecules (the number of such molecules is negligibly small).  In the computation of AAD of the tetrahedral network, we consider N in Eq.~(\ref{eq:aad}) as the number of molecules constituting the network.   

\section{Results and discussion}

\begin{figure}
\includegraphics[width=0.5\textwidth]{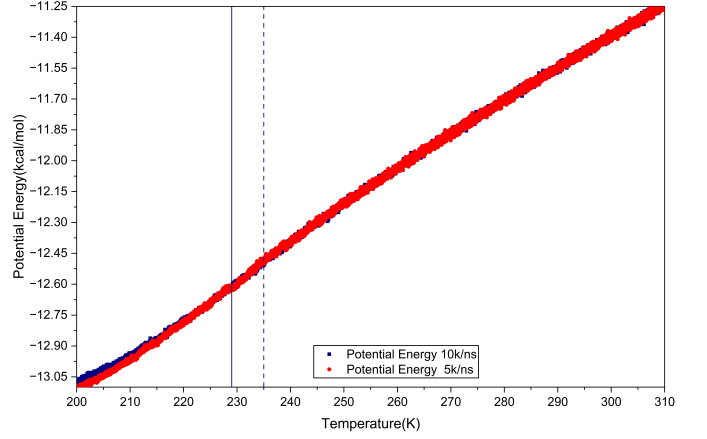}% Here is how to import EPS art
\caption{Instantaneous potential energy values (in kcal/mol) sampled after even 2 ps as a function of temperature in the 5 K/ns and 10 K/ns trajectories.  The dashed and solid vertical lines (in this and all subsequent figures)  correspond to `dynamical crossover' point and `freezing' point in the 5 K/ns trajectory.}
\label{fig:pe}
\end{figure}

\begin{figure}
\includegraphics[width=0.5\textwidth]{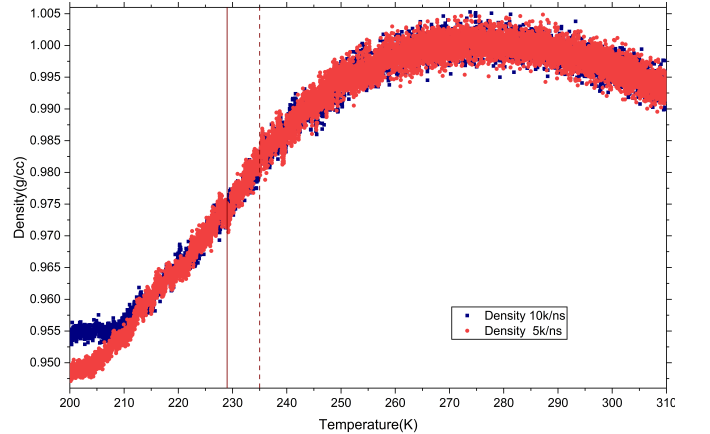}% Here is how to import EPS art
\caption{Instantaneous density values (in g/cc) sampled after even 2 ps as a function of temperature in the 5 and 10 K/ns trajectories.}
\label{fig:density}
\end{figure}

In Figs.~\ref{fig:pe} and \ref{fig:density}, the instantaneous values of the potential energy and density from the cooling trajectories of 5 K/ns and 10 K / ns are plotted as functions of temperature.   The vertical lines demarcate the `transition region' between the ‘dynamical crossover’ point~\cite{GAUTAM17,PINGUA18} (dashed line) and the ‘freezing’ point  of the network (solid line).  These points are defined with respect to the slower trajectory.  Both trajectories correctly reproduce the density maximum at $T_0 = 277$ K as well as the average value of the potential energy at the freezing point ($T_m = 252.1$ K).~\cite{ABASCAL05}   As the cooling continues in the supercooled liquid region, the averaged potential energy and density from the two trajectories continue to agree upto the dynamical crossover point (dashed vertical line), which signals the end of the liquid state.  Upon further cooling,  the two trajectories start to deviate,  with the slower (5 K/ns) trajectory exhibiting lower potential energy.    This indicates higher degree of structural relaxation in the 5 K/ns trajectory.    We note that a qualitatively similar behavior is seen in the cooling trajectories of water modeled with the mW potential by Moore and Molinero.~\cite{MOORE11-2} In their results, the 1 K/ns trajectory shows a systematically lower enthalpy as compared to 10 K/ns trajectory across the limit of stability of the mW water (see Fig.~1 of Ref.~\onlinecite{MOORE11-2}).    This occurs due to higher degree of crystallization (structural transformation) in the slower trajectory.~\cite{MOORE11-2} Although the deviation between the two trajectories in mW water is much larger (as compared to our results), we expect a similar phenomenon of structurally driven transition as the underlying cause of the deviation.  Henceforth, we will elucidate on the changes  in the slower trajectory due to its higher degree of relaxation.
Figure~S1 (see Supplementary Information file) shows the zoomed-in portion of  the potential energy changes across  the transition region in the 5 K/ns trajectory.   After the dynamical crossover, large scale thermal fluctuations in the potential energy dissipate while the thermal equilibration is achieved due to small scale fluctuations.   This marks the end of the supercooled liquid state.~\cite{GAUTAM17}  Subsequently, the potential energy (averaged over small scale fluctuations) decreases continuously as the cooling continues and achieves a stationary condition (an inflection point) at the freezing point of the tetrahedral network.   We note that a similar behavior is seen across the dynamical crossover point of water simulated with the mW potential at its limit of stability :~\cite{PINGUA18}  a sharp decrease in the potential energy after the dynamical crossover (vertical line in Fig. 3(b) of Ref.~\onlinecite{PINGUA18}) followed by stationary condition of potential energy which is an inflection point [at around $45 \times 10^4$ MD step in Fig. 3(b) of Ref.~\onlinecite{PINGUA18}].

\begin{figure}
\includegraphics[width=0.5\textwidth]{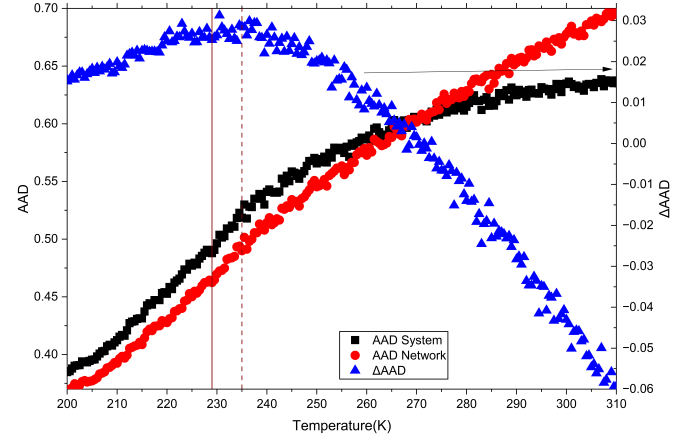}% Here is how to import EPS art
\caption{\label{fig:aad} AAD of the system and the AAD of the network [see Eq.~(\ref{eq:aad})] as a function of temperature in the 5 K/ns trajectory.  The right hand
ordinate refers to  $\Delta$AAD which is the AAD of the system minus AAD of the network}
\end{figure}

\begin{figure}
\includegraphics[width=0.5\textwidth]{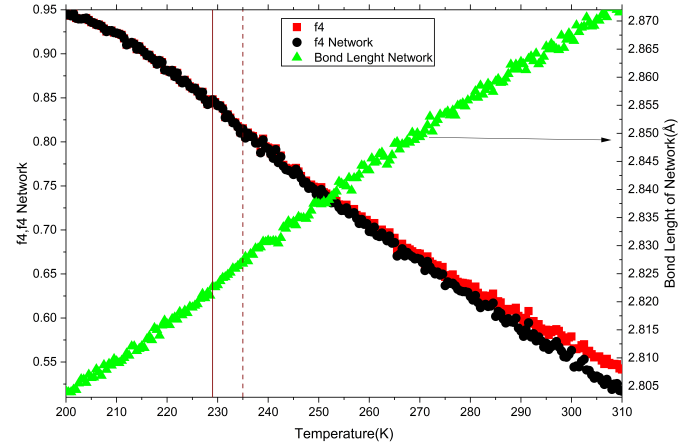}% Here is how to import EPS art
\caption{\label{fig:f4} The fraction of 4-coordinated molecules in the system (f4) and that
of the network (f4c) and the average bond length of the network (right ordinate) as a function of temperature in the 5 K/ns trajectory.}
\end{figure}

To analyze the corresponding network related changes in the 5 K/ns trajectory, we have plotted the AAD of the entire system as well as that of the network in Fig.~\ref{fig:aad}.  Also plotted is the difference between the two AAD curves.   As expected, the AAD of the system decreases continuously as the water is cooled across the temperature of maximum density, showing an increase in the tetrahedrality throughout the system. However, the AAD of the network decreases at a faster rate  (which can be seen clearly from the difference plot) due to the inherently cooperative nature of the hydrogen bonded network.~\cite{STILLINGER80} Across the dynamical crossover, the large scale thermal fluctuations in the AAD of the system and the AAD of the network dissipate (see the zoomed-in version of the AAD plots in Fig.~S3 of the Supplementary information file).   Subsequently, we find that there is a monotonic decrease in the AAD of the network .  To analyze the related changes in the bond lengths and the size of the network, we have plotted the fraction of 4 coordinated molecules constituting the network and the fraction of total number of 4 coordinated molecules in the system,  along with the average bond length of the network in Fig.~\ref{fig:f4}.  It can be seen that there is an overall decrease in the average bond length of the network as the network grows in size, i.e., f4c increases (although thermal fluctuations persist in both quantities) during the cooling process.  As the cooling approaches the dynamical crossover point, most of the 4-coordinated molecules in the system merge with the network (f4c and f4 curves collapse).      The fluctuations in the average bond length of the network  dissipate at the dynamical crossover point and subsequently average bond length decreases continuously  and smoothly (see the zoomed-in version in Fig.~S4 of Supplementary information file), while the size of the network increases. Thus there is a continuous tightening of the bonds of the network which is a distinctive feature of the transition region. 

Through the simultaneous decrease in the AAD (indicating bond angle relaxation) as well as the average bond length, the network, as a whole, drives the decrease of the potential energy of the entire system past the dynamical crossover.    This results in an acceleration in the rate of decrease of energy in the transition region as seen from Fig.~\ref{fig:pe}. This process of network driven relaxation  terminates when the network undergoes `freezing’ (at the solid vertical line), i.e., the average bond length of the network achieves approaches a stationary condition (see Fig.~S4 of Supplementary Information file) indicating that network develops rigidity, due to which the network evolution, as a whole, can no longer continue.   As a result, the potential energy of the system also approaches a stationary condition at the solid vertical line (which is an inflection point as seen Fig.~S1 of Supplementary Information file).  Due to freezing of the network, the system goes through a mechanical discontinuity across the solid vertical line:  the (numerical) derivatives of the density and the network size with respect to energy (when all quantities are averaged over the small scale fluctuations) are discontinuous across the freezing point, i.e., the density increases and the size of the network decreases even as the energy decreases past the freezing point (see the zoomed-in Figs. S1, S2 and S4 of Supplementary Information file).  In an earlier work, a similar discontinuity is seen in the network size across the inflection point of potential energy in water simulated using the mW potential [see the changes at around $45 \times 10^4$ MD step in Figs. 3(b) and 5(b) of Ref.~\onlinecite{PINGUA18}].   Interestingly, the results of Matsumoto also indicate an apparent discontinuity in the specific volume, the fraction of 4-coordinated molecules, as well as the volumes of Voronoi Polyhedra of 3, 4, and 5-coordinated molecules at around 230 K in the supercooled region (see Figs.~1 and 2 of Ref.~\onlinecite{mMATSUMOTO09}).  Note that this temperature (230 K) is close to the temperature at which freezing of the network has been observed in our results.  The fraction of 4-coordinated molecules (see the value of C4 in upper panel of Fig.~2 of Ref.~\onlinecite{mMATSUMOTO09}) is about 83 \% at around 230 K, which closely agrees with the value of f4 $\approx$ 85 \% at the freezing point (see the value at the solid vertical line in  Fig.~S4 of Supplementary Information file).

We find that the mechanical discontinuity is associated with a change in the relaxation mechanism induced by the freezing of the network :  the subsequent relaxation occurs due to an increase in the structural order (crystallization) .     It can be seen that across the freezing point, the average difference between the AAD of the system and the network approaches a  stationary condition (see the $\Delta$AAD plot in Fig.~\ref{fig:aad}) and thereafter, the average AAD difference decreases as the cooling continues.  This indicates that the system-wide angular relaxation (due to structural ordering) drives the decrease of the AAD of the network.  Thus, our results indicate that the relaxation occurs through  cooperative ordering or cooperative crystallization past the freezing point.   It is interesting to note that in the Ni3Fe alloy system a similar phenomenon has been found: the freezing of the L12 ordered domains induces cooperative crystallization across the critical cooling temperature of the Ni3Fe alloy.~\cite{MANGLA22}

\begin{figure}
\includegraphics[width=0.5\textwidth]{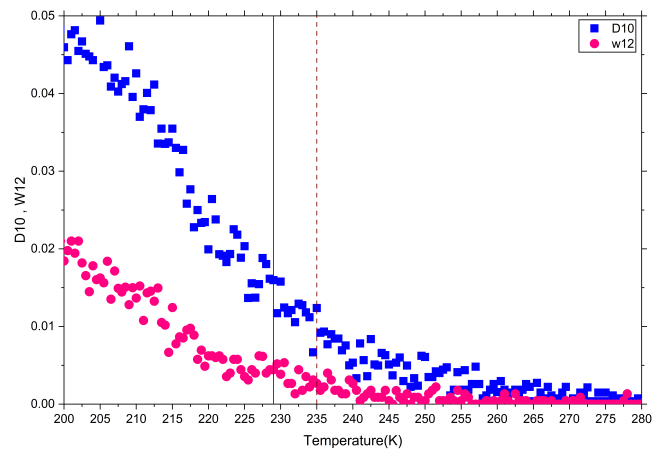}% Here is how to import EPS art
\caption{\label{fig:d10w12} Fractions of molecules in the cubic (D10) and hexagonal (W12) structural motifs as a function of temperature in the 5 K/ns trajectory.}
\end{figure}

\begin{figure}
\includegraphics[width=0.5\textwidth]{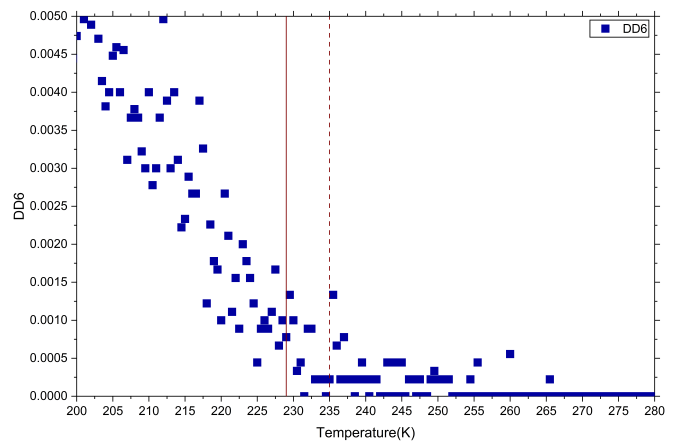}% Here is how to import EPS art
\caption{\label{fig:dd6}Fraction of DD6 molecules as a function of temperature in the 5 K/ns trajectory.}
\end{figure}

In what follows, we elucidate the key structural changes past the freezing point.  We consider the MRO motifs (vitrites) found in cubic and hexagonal crystalline phases.   In particular, we trace the diamond (D10) units formed by 4 chair rings and the wurtzite units (W12) formed by 2 chair and 3 boat rings (see Fig. 1 of Ref.~\onlinecite{mMATSUMOTO08} and Figs.~2(d) and 2(h) of Ref.~\onlinecite{PINGUA18}).  Note that D10 units and W12 units were designated as `D' and `W' in Ref.~\onlinecite{PINGUA18}.  In Fig.~\ref{fig:d10w12}, we have plotted the fraction of molecules of D10 and W12 units.  We find that the fractions of both D10 and W12 molecules increase at an accelerating pace  after the freezing point and the rate of increase is maximized at around 216 K simultaneously.   As the D10 units increase, some of these units join together through the shared chair rings.  The molecules of the shared chair rings are labeled as DD6 molecules [see Fig.~2(e) of Ref.~\onlinecite{PINGUA18}].   In Fig.~\ref{fig:dd6}, it can be seen that the fraction of DD6 molecules increases at an accelerated pace below the freezing point of the network.   Although the overall extent of DD6 molecules is low, our results point to an important aspect of structural ordering :  the D10 units tend to join together leading to hierarchically ordered structures in cubic ice.    We measure the overall extent of crystallization as the percentage of total number of D10 and W12 molecules.  Thus the overall extent of crystallization is around 6.5\% at 200 K in the LDA ice while the extent of hexagonal ice (i.e., the percentage of W12 molecules) is around 2\%.  Thus, the fraction of D10 molecules is significantly higher compared to the fraction of W12 molecules. These results can be refined further  using the specific criteria~\cite{PINGUA18, PINGUA19} for identifying the boat shaped and chair shaped rings that constitute the D10 and W12 units.  We also note that the overall extent of crystallization as well as relative extents of cubic and hexagonal ices in the simulated LDA states depend upon the molecular model used.   Nonetheless, our results are consistent with the experimental fact that cubic ice forms a  significant fraction of total ice in the LDA states.~\cite{AMAYA17}     

\section{Conclusions}

The main contribution of this work is that it provides the specific mechanism by which crystallization is initiated across the limit of stability of supercooled water.    Our work shows that the freezing of the tetrahedral amorphous network  triggers the crystallization across the limit of stability of supercooled water modeled with the TIP4P/2005 potential.  The freezing of the tetrahedral network is preceded by the dynamical crossover of thermal fluctuations and subsequent network driven relaxation in the transition region (demarcated by two vertical lines in our results).   A similar relaxation mechanism has been found earlier at the temperature corresponding to limit of stability of water simulated with the monatomic (mW) potential.~\cite{PINGUA18} Although the extent of crystallization may depend on the specific potential used, qualitatively similar nature of the underlying phenomena found in  two independent potentials of water indicates that the mechanism may be applicable for real water as well.   In an earlier work, the freezing of the tetrahedral amorphous network was also found to be relevant in the relaxation of supercooled silicon~\cite{APTE15} modeled by the Stillinger-Weber potential.~\cite{STILLINGER85}   This indicates that the crystallization mechanism explored in this work may be applicable to a wider class of supercooled tetrahedral liquids. This needs to be confirmed further by using more realistic force fields for water as well as other tetrahedral liquids such as silicon.  We plan to conduct such studies as a follow-up of this work.
%%%%%%%%%%%%%%%%%%%%%%%%%%%%%%%%%%%%%%%%%%%%%%%%%%%%%%%%%%%%%%%%%%%%%%%%%%%%%%%%%%%%%%%%%%%%%%%%%%%%%%%%%%

\begin{acknowledgments}
P. A. gratefully acknowledges the late Dr. B. D. Kulkarni (CSIR-National Chemical Laboratory) for his insightful comments and guidance on the network freezing phenomenon.  The simulations reported here were performed on the PARAM Sanganak supercomputer at the Indian Institute of Technology Kanpur, established under the National Supercomputing Mission (NSM), Government of India, and supported by the Department of Science and Technology (DST) and Ministry of Electronics and Information Technology (MeitY).
\end{acknowledgments}

\nocite{}
\bibliography{aipsamp}% Produces the bibliography via BibTeX.

@article{ABASCAL05,
  author  = {J. L. F. Abascal and C. Vega},
  title   = {{A general purpose model for the condensed phases of water: TIP4P/2005}},
  journal = {J. Chem. Phys.},
  volume  = 123,
  pages   = {234505},
  year    = 2005
}

@article{AMAYA17,
  author  = {Andrew J. Amaya and Harshad Pathak and Viraj P. Modak 
             and Hartawan Laksmono and N. Duane Loh and Jonas A. Sellberg
          and Raymond G. Sierra and Trevor A. McQueen and Matt J. Hayes
           and Garth J. Williams and Marc Messerschmidt and Sebastien
           Boutet and Michael J. Bogan and Anders Nilsson 
           and Claudiu A. Stan and Barbara E. Wyslouzil},
  title   = {{How cubic can ice be ?}},
  journal = {J. Phys. Chem. Lett.},
  volume  = 8,
  pages   = {3216--3222},
  year    = 2017
}

@article{APTE15,
  author  = {Pankaj A. Apte and Nandlal Pingua and Arvind K. Gautam and Uday Kumar
             and Soohaeng Yoo Willow and Xiao Cheng Zeng and B. D. Kulkarni},
  title   = {{The freezing tendency towards 4-coordinated amorphous networks causes an increase
              in heat capacity of supercooled Stillinger--Weber silicon}},
  journal   = {RSC Adv.},
  volume  = 5,
  pages   = {44679--44686},
  year    = 2015
}

@article{BENMORE05,
  author  = {C. J. Benmore and R. T. Hart and Q. Mei and D. L. Price and J. Yarger and C. A. Tulk and D. D. Klug},
  title   = {{Intermediate Range Chemical Ordering in amorphous and liquid water, Si, and Ge}},
  journal = {J. Phys. Chem. B},
  volume  = 72,
  pages   = {132201},
  year    = 2005
}

@article{ELLIOTT95,
  author  = {S. R. Elliott},
  title   = {{Interpretation of the principal diffraction peak of liquid  and amorphous water}},
  journal = {J. Chem. Phys.},
  volume  = 103,
  pages   = {2758--2761},
  year    = 1995
}

@article{GAUTAM17,
  author  = {Arvind Kumar Gautam and Nandlal Pingua and Aashish Goyal and Pankaj A. Apte},
  title   = {{Dynamical instability causes the demise of a supercooled tetrahedral liquid}},
  journal = {J. Stat. Phys.},
  volume  = 168,
  pages   = {1302--1318},
  year    = 2017
}

@article{HUJO11,
  author  = {Waldmer Hujo and B. Shadrack Jabes and Varun K. Rana and Charusita
            Chakravarti and Valeria Molinero},
  title   = {{The rise and fall of anamolies in tetrhedral liquids}},
  journal = {J. Stat. Phys.},
  volume  = 145,
  pages   = {293--312},
  year    = 2011
}

@article{LUZAR93,
  author  = {Alenka Luzar and David Chandler},
  title   = {{Structure and hydrogen bond dynamics of water-dimethyl sulfoxide
mixtures by computer simulations}},
  journal = {J. Chem. Phys.},
  volume  = 98,
  pages   = {8160--8173},
  year    = 1993
}

@article{LUZAR96,
  author  = {Alenka Luzar and David Chandler},
  title   = {{Hydrogen-bond kinetics in liquid
water}},
  journal = {Nature},
  volume  = 379,
  pages   = {55--57},
  year    = 1996
}

@article{MALKIN12,
  author  = {Tamsin L. Malkin and Benjamin J. Murray and Andrey V. Brukhno and 
             Jamshed Anwar and Christoph G. Salzmann},
  title   = {{Structure of ice crystallized from supercooled water}},
  journal = {Proc Natl Acad Sci USA},
  volume  = 109,
  pages   = {1041--1045},
  year    = 2012
}

@article{MALKIN15,
  author  = {Tamsin L. Malkin and Benjamin J. Murray and Christoph G. Salzmann
             and Valeriz Molinero and Steven J. Pickering and Thomas F. Whale},
  title   = {{Stacking disorder in ice I}},
  journal = {Phys. Chem. Chem. Phys.},
  volume  = 17,
  pages   = {60--76},
  year    = 2015
}

@article{MANGLA22,
  author  = {Anil Mangla and Goutam Deo and Pankaj A. Apte},
  title   = {{Cooperative freezing of the L12 ordered domains at the critical cooling temperature of Ni3Fe alloy}},
  journal = {J. Stat. Mech.},
  pages   = {093204},
  year    = 2022
}

@article{mMATSUMOTO02,
  author  = {Masakazu Matsumoto and Shinji Saito and Iwao Ohmine},
  title   = {{Molecular dynamics simulation of the ice nucleation and growth
             process leading to water freezing}},
  journal = {Nature},
  volume  = 416,
  pages   = {409--413},
  year    = 2002
}

@article{mMATSUMOTO08,
  author  = {Masakazu Matsumoto and Akinori Baba and Iwao Ohmine},
  title   = {{Network motif of water}},
  journal = {AIP Conf. Proc.},
  volume  = 982,
  pages   = {219--222},
  year    = 2008
}

@article{mMATSUMOTO09,
  author  = {Masakazu Matsumoto},
  title   = {{Why does water expand when it cools?}},
  journal = {Phys. Rev. Lett.},
  volume  = 103,
  pages   = {017801},
  year    = 2009
}

@article{MAYER87,
  author  = {Erwin Mayer and Andreas Hallbrucker},
  title   = {{Cubic ice from liquid water}},
  journal = {Nature},
  volume  = 325,
  pages   = {601--602},
  year    = 1987
}

@article{MOLINERO09,
  author  = {Valeria Molinero and Emily B. Moore},
  title   = {{Water modeled as an intermediate element between Carbon and Silicon}},
  journal = {J. Phys. Chem. B},
  volume  = 113,
  pages   = {4008--4016},
  year    = 2009
}

@article{MOORE10,
  author  = {Emily B. Moore and Valeria Molinero},
  title   = {{Ice crystallization in water's ``no man's land''}},
  journal = {J. Chem. Phys.},
  volume  = 132,
  pages   = {244504},
  year    = 2010
}

@article{MOORE11-1,
  author  = {Emily B. Moore and Valeria Molinero},
  title   = {{It it cubic ? Ice crystallization from deeply supercooled water}},
  journal = {Phys. Chem. Chem. Phys.},
  volume  = 13,
  pages   = {20008-20016},
  year    = 2011
}

@article{MOORE11-2,
  author  = {Emily B. Moore and Valeria Molinero},
  title   = {{Structural transformation in supercooled water controls the
              crystallization rate of ice}},
  journal = {Nature},
  volume  = 479,
  pages   = {506--509},
  year    = 2011
}

@article{MURRAY06,
  author  = {Benjamin J. Murray and Allan K. Bertram},
  title   = {{Formation and stability of cubic ice in water droplets}},
  journal = {Phys. Chem. Chem. Phys.},
  volume  = 8,
  pages   = {186--192},
  year    = 2006
}

@article{PINGUA18,
  author  = {Nandlal Pingua and Pankaj A. Apte},
  title   = {{Increase in local tetrahedral order across the limit of stability
             leads to cubic-hexagonal stacking in supercooled monatomic (mW) water}},
  journal = {J. Chem. Phys.},
  volume  = 149,
  pages   = {074506},
  year    = 2018
}

@article{PINGUA19,
  author  = {Nandlal Pingua and Pankaj A. Apte},
  title   = {{Topological identification criteria, stability, and relevance of
pentagonal nanochannels in amorphous ice}},
  journal = {J. Phys. Chem. B},
  volume  = 123,
  pages   = {10301--10310},
  year    = 2019
}

@article{PLIMPTON95,
  author  = {S. Plimpton},
  title   = {{Fast parallel algorithms for short--range
             molecular dynamics}},
  journal = {J. Comp. Phys.},
  volume  = 117,
  pages   = {1--19},
  year    = 1995
}

@article{STILLINGER80,
  author  = {Frank H. Stillinger},
  title   = {{Water Revisited}},
  journal = {Science},
  volume  = 209,
  pages   = {451--457},
  year    = 1980,
}

@article{STILLINGER85,
  author  = {Frank H. Stillinger and Thomas A. Weber},
  title   = {{Computer Simulation of Local Order in Condensed Phases of Silicon}},
  journal = {Phys. Rev. B},
  volume  = 31,
  pages   = {5262--5271},
  year    = 1985,
}

\end{document}